\documentstyle[12pt,epsf,rotating,axodraw]{article}

\catcode`@=11
\def\citer{\@ifnextchar
[{\@tempswatrue\@citexr}{\@tempswafalse\@citexr[]}}

\def\@citexr[#1]#2{\if@filesw\immediate\write\@auxout{\string\citation{#2}}\fi
  \def\@citea{}\@cite{\@for\@citeb:=#2\do
    {\@citea\def\@citea{--\penalty\@m}\@ifundefined
       {b@\@citeb}{{\bf ?}\@warning
       {Citation `\@citeb' on page \thepage \space undefined}}%
\hbox{\csname b@\@citeb\endcsname}}}{#1}}
\catcode`@=12

\newcommand{\gsim}{\raisebox{-0.13cm}{~\shortstack{$>$ \\[-0.07cm] $\sim$}}~}

\newcommand{\MS}{\overline{\rm MS}}
\newcommand{\non}{\nonumber}
\newcommand{\beq}{\begin{equation}}
\newcommand{\eeq}{\end{equation}}
\newcommand{\beqn}{\begin{eqnarray}}
\newcommand{\eeqn}{\end{eqnarray}}

\begin{document}

\renewcommand{\thefootnote}{\fnsymbol{footnote}}

\begin{flushright}
DESY 02--035 \\
KA--TP--34--2001 \\
PSI--PR--01--15 \\
hep-ph/0203269 \\[1.0cm]
\end{flushright}

\begin{center}
{\Large \bf Photoproduction of $W$ Bosons at HERA: \\[0.5cm]
                   QCD Corrections\footnote{This work has been supported
in part by the Swiss Bundesamt f\"ur Bildung und Wissenschaft, by the
European Union under contract HPRN-CT-2000-00149 and by the
DFG Research Group 'Quantenfeldtheorie, Computeralgebra und
Monte-Carlo-Simulation'.}
} \\[1.0cm]

{\sc Kai--Peer O.\,Diener$^1$, Christian Schwanenberger$^{2,3}$ and
Michael Spira$^2$} \\[1.0cm]
{\it $^1$ Institut f\"ur Theoretische Physik, Universit\"at Karlsruhe,
D--76128 Karlsruhe, Germany \\
$^2$ Paul Scherrer Institut PSI, CH--5232 Villigen PSI, Switzerland \\
$^3$ Deutsches Elektronen-Synchrotron DESY, D--22603 Hamburg,
Germany\footnote{Present address}} \\[1.0cm]
\end{center}

\begin{abstract}
\noindent
$W$ bosons can be produced in the channels $e^\pm p\to W^\pm + X$ at
HERA thus allowing to probe for anomalous trilinear couplings among the
gauge bosons.  We discuss the next-to-leading order (NLO) QCD
corrections to the photoproduction of $W$ bosons
with finite transverse momentum at HERA. The higher-order QCD corrections
reduce the factorization scale dependence significantly and modify the
leading-order (LO) cross sections by $\pm {\cal O}(10\%)$.
\end{abstract}

\renewcommand{\thefootnote}{\arabic{footnote}}
\setcounter{footnote}{0}

\section{Introduction}
The study of $W$ bosons at different colliders serves as an important test
of the Standard Model and possible extensions. $W$ bosons can be
produced at the $ep$ collider HERA with a center of mass (c.m.)
energy $\sqrt{s}\approx 318$ GeV which is achieved by colliding
electrons/positrons with energy $E_e=27.5$ GeV and protons with energy
$E_p=920$ GeV. Since the production cross sections for the
processes $e^\pm p \to e^\pm W + X$ reach values of about 1 $pb$ at HERA,
the production mechanisms of $W$ bosons can be studied and the existence
of anomalous $WW\gamma$ trilinear couplings can be probed
\citer{heratri,dubinin}. Moreover, $W$ boson
production represents an important SM background to several new physics
searches. In particular it is the dominant SM process leading to isolated
high energy lepton events with missing transverse momentum
\cite{heratri,muon}. In order to determine potential discrepancies between
measurements and Standard Model (SM) predictions, the latter have to be
sufficiently accurate and reliable. This is not guaranteed for the
available LO calculations of $W$ boson production
\cite{baur,dubinin,lo}. For an unambiguous test of anomalous contributions,
it is necessary to extend the previous analyses to NLO accuracy. The first
step in this direction has been made in Ref.~\cite{NRS}, where the QCD
corrections to the total resolved photoproduction cross section have been
determined. However, the result cannot be used for $W$ boson production
with large transverse momentum which is dominated
by direct photoproduction, the QCD corrections to which are discussed in this
letter.

\section{Leading Order}
$W$ boson production at $ep$ colliders is mediated by photon, $Z$
and $W$ exchange between the electron/positron and the hadronic currents of
the process. In general two phase-space regions are distinguished:
the deep inelastic (DIS) regime at
large $Q^2$ and the photoproduction regime at small $Q^2$, $Q^2$ being the
negative square of the transferred momentum from the electron/positron. The
photoproduction cross section can
be calculated by convoluting the Weizs\"acker-Williams photon spectrum,
\beq
f_{\gamma/e}(x_\gamma) = \frac{\alpha}{2\pi} \left\{
\frac{1+(1-x_\gamma)^2}{x_\gamma} \log
\frac{Q^2_{max}}{Q^2_{min}} - 2 m_e^2 x_\gamma \left( \frac{1}{Q^2_{min}} -
\frac{1}{Q^2_{max}} \right) \right\},
\label{eq:wwa}
\eeq
with the differential hadronic cross section with respect to the
transverse momentum $p_{TW}$ and the rapidity $y_W$ of the $W$ boson for
$\gamma q \to q' W$:
\beq
\frac{d^2\sigma^{dir}_{LO}}{dp_{TW} dy_W} = 2p_{TW} \sum_{q,\bar q}
\int_{x_\gamma^-}^1 \frac{dx_\gamma}{x_\gamma}
f_{\gamma/e} (x_\gamma) q_p(x_p,\mu_F^2) \frac{x_p}{s+u_1}
\frac{s^2}{S} \frac{d \hat \sigma^{dir}_{LO}(\gamma q\to Wq')}{dt_1} ,
\label{eq:dsiglo}
\eeq
where $q_p(x_p,\mu_F^2)$ is the corresponding quark density of the proton
with momentum fraction $x_p$ and at the factorization scale $\mu_F$.
$\hat \sigma^{dir}_{LO}(\gamma q\to Wq')$ denotes the corresponding LO
partonic cross section and $x_\gamma$ the photon momentum fraction of
the electron/positron.
(For the resolved part the Weizs\"acker--Williams spectrum $f_{\gamma/e}$
has to be replaced by the convolution of $f_{\gamma/e}$ with
the corresponding parton densities in the photon and the partonic cross
section by the resolved expressions.) The QED coupling at vanishing
momentum transfer is denoted by $\alpha$, the electron mass by $m_e$, and
the minimal and maximal values of the photon virtuality $Q^2$
by $Q^2_{min}, Q^2_{max}$.
In order to separate photoproduction from the DIS region we
impose a cut of $Q^2_{max} = 4~{\rm Ge\!V}^2$ in the momentum transfer
$Q^2$ from the electron/positron line. The minimal value of $Q^2$ is
kinematically fixed,
\beq
Q^2_{min} = m_e^2 \frac{x_\gamma^2}{1-x_\gamma},
\eeq
where negligible higher order terms in the electron mass $m_e$ have been
omitted. The partonic Mandelstam variables are defined as
\beqn
s & = & (k_1+k_2)^2, \non \\
t & = & (k_1-p_1)^2, \hspace*{1cm} t_1 = t - M_W^2, \non \\
u & = & (k_2-p_1)^2, \hspace*{1cm} u_1 = u - M_W^2,
\label{eq:mandelpart}
\eeqn
where $k_1$ is the 4-momentum of the photon, $k_2$ the incoming quark
4-momentum and $p_1$ the 4-momentum of the outgoing $W$ boson. The
momentum fraction $x_p$ of the proton is given by $x_p = x_p^-$, where
\beq
x_p^- = -\frac{x_\gamma T_1 + M_W^2}{x_\gamma S + U_1},
\eeq
and the lower bound of the $x_\gamma$ integration by
\beq
x_\gamma^- = -\frac{U_1 + M_W^2}{S + T_1}.
\eeq
$S,T_1,U_1$ denote the hadronic Mandelstam variables. They are related to the
partonic variables as
\beq
s=x_\gamma x_pS, \qquad t_1 = x_\gamma T_1, \qquad u_1 = x_p U_1.
\eeq
The hadronic Mandelstam variables can be expressed in terms of the transverse
momentum $p_{TW}$ and the rapidity $y_W$ (defined to be positive in the
electron/positron direction) of the $W$ boson,
\beq
T_1 = -\sqrt{S(p_{TW}^2+M_W^2)} e^{-y_W-y_0}, \qquad
U_1 = -\sqrt{S(p_{TW}^2+M_W^2)} e^{y_W+y_0},
\label{eq:mandelhad}
\eeq
where $y_0=\frac{1}{2} \log E_p/E_e$ denotes the shift in rapidity
between the laboratory system and the hadronic c.m.~system.

\begin{figure}[hbtp]
\begin{center}
\begin{picture}(120,70)(20,10)

\ArrowLine(0,20)(50,20)
\ArrowLine(50,20)(50,80)
\ArrowLine(50,80)(100,80)
\Photon(50,20)(100,20){3}{5}
\Photon(0,80)(50,80){3}{5}

\put(-15,78){$\gamma$}
\put(-15,18){$q$}
\put(55,48){$q'$}
\put(105,78){$q'$}
\put(105,18){$W$}

\end{picture}
\begin{picture}(120,70)(-10,10)

\ArrowLine(0,80)(50,80)
\ArrowLine(50,80)(50,20)
\ArrowLine(50,20)(0,20)
\Gluon(50,80)(100,80){3}{5}
\Photon(50,20)(100,20){3}{5}

\put(-15,78){$q$}
\put(-15,18){$\bar q´$}
\put(55,48){$q$}
\put(105,78){$g$}
\put(105,18){$W$}

\end{picture}
\begin{picture}(120,70)(-40,10)

\ArrowLine(0,80)(50,80)
\ArrowLine(50,80)(100,80)
\ArrowLine(0,20)(50,20)
\ArrowLine(50,20)(50,50)
\ArrowLine(50,50)(100,50)
\Photon(50,20)(100,20){3}{5}
\Photon(50,50)(50,80){3}{3}

\put(-15,78){$e$}
\put(-15,18){$q$}
\put(55,33){$q'$}
\put(55,63){$\gamma,Z$}
\put(105,78){$e$}
\put(105,48){$q'$}
\put(105,18){$W$}

\end{picture}  \\
\caption[]{\label{fg:dia} \it Typical diagrams of $W$ boson production with
finite transverse momentum at HERA: direct, resolved and DIS mechanism.}
\end{center}
\vspace*{-0.5cm}
\end{figure}
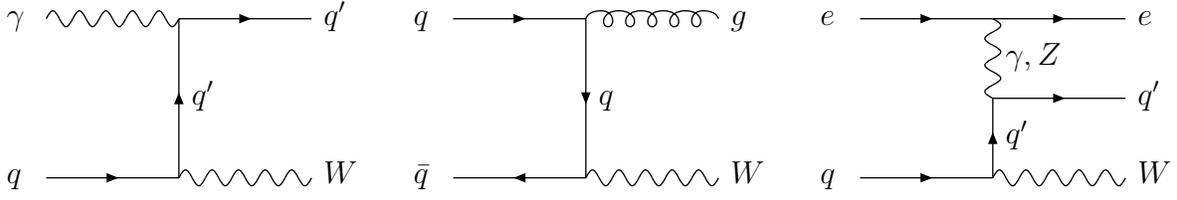

The leading direct photon process $\gamma q\to q'W$ (a typical
contribution is depicted by the first diagram of Fig.~\ref{fg:dia})
develops a singularity at LO when the final state quark $q'$ becomes collinear
with the initial state photon. However, the finite transverse momentum
$p_{TW}$ of the $W$ boson has to be balanced by the final state quark, so
that this singularity does not occur at LO for non-vanishing
$p_{TW}$. The small $Q^2$ region includes the contribution of
the hadronic component of the photon giving rise to $W+jet$ production
via $q\bar q'\to Wg$ (second diagram of Fig.~\ref{fg:dia} as a typical
example) and the crossed processes $gq(\bar q) \to W q'(\bar q')$.
The treatment of the DIS region is straightforward (a typical
contribution is shown in the third diagram of Fig.~\ref{fg:dia}).

The result for the direct contribution can be cast into the form
\beq
\frac{d\hat \sigma^{dir}_{LO}}{dt_1} [\gamma q \to Wq'] =
-\frac{G_F M_W^2\alpha}{\sqrt{2}}
\frac{(e_{q'}t_1-e_Wu)^2}{s^2 u t_1^2} \left( s^2+u^2+2M_W^2t \right) ,
\eeq
where $G_F$ denotes the Fermi constant, $M_W$ the $W$ mass and $e_{q'},
e_W$ the electric charges of the scattered quark $q'$ and $W$ boson, i.e.\
$e_W = e_q-e_{q'} = \pm 1$. The LO cross sections of the
resolved processes $q \bar q'\to Wg$ and $gq(\bar q) \to W q(\bar q)$ are
given by ($C_F=4/3$ and $T_R=1/2$)
\beqn
\frac{d\hat \sigma^{res}_{LO}}{dt_1} [q\bar q' \to Wg] & = &
\frac{G_F M_W^2\alpha_s}{3\sqrt{2}} C_F \frac{t^2+u^2+2M_W^2s}{s^2 tu}, \non \\
\frac{d\hat \sigma^{res}_{LO}}{dt_1} [gq(\bar q) \to Wq'(\bar q')] & = &
- \frac{G_F M_W^2\alpha_s}{3\sqrt{2}} T_R \frac{s^2+u^2+2M_W^2t}{s^3 u},
\eeqn
where $\alpha_s$ denotes the strong coupling constant.
\begin{figure}[hbtp]
\vspace*{0.5cm}

\hspace*{2.0cm}
\begin{turn}{-90}%
\epsfxsize=8cm \epsfbox{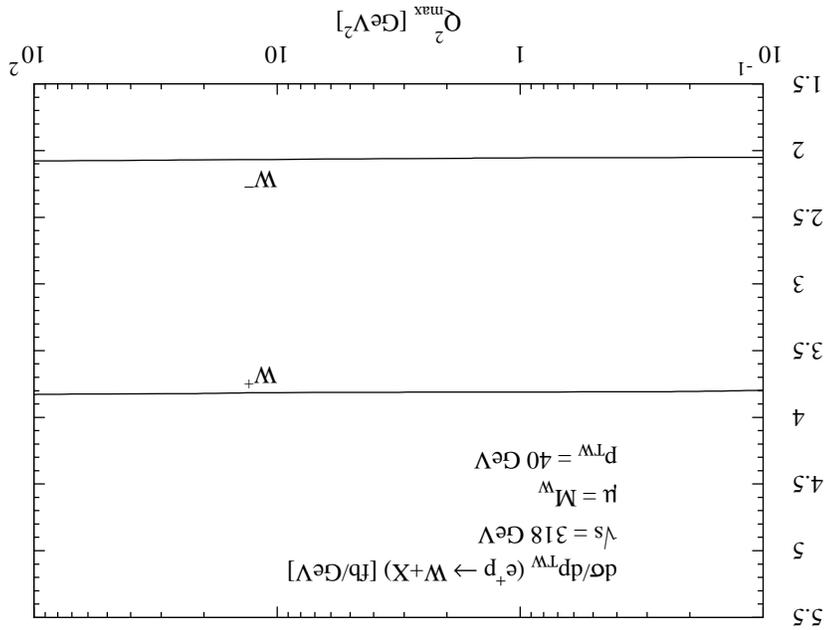}

\end{turn}
\vspace*{0.0cm}

\caption[]{\it \label{fg:q2} Dependence of the differential LO
$W^\pm$ production cross sections for $p_{TW}=40$ GeV on the cut $Q^2_{max}$
which separates the DIS and photoproduction regimes, after adding the DIS,
direct and resolved contributions. The renormalization and factorization
scales have been identified with the $W$ mass, $\mu=\mu_F=\mu_R=M_W$.}
\end{figure}

The direct, resolved and DIS contributions add up to the total $W$
$p_{TW}$ distribution. Direct photoproduction develops the dominant
contribution, while the DIS part is smaller but competitive. The
resolved component is negligible for $p_{TW} \gsim 15$ GeV \cite{spira}.
The consistency of the calculation requires that
the dependence on the specific value of the cut $Q^2_{max}$ which
separates the DIS and photoproduction regimes, should be small. This
dependence is presented in Fig.~\ref{fg:q2} for the LO $W^+$ and $W^-$
cross sections at $p_{TW}=40$ GeV. The residual
dependence is below the per-cent level and thus indeed sufficiently small.

\section{QCD Corrections}
For the dominant direct part we have evaluated the NLO QCD corrections.
They consist of two parts, the
virtual and real corrections. The virtual corrections are built up by all
one-loop diagrams which are generated by virtual gluon exchange.
Typical examples are displayed in Fig.~\ref{fg:diavirt}. The real corrections
originate from gluon radiation off the quark lines, see
Fig.~\ref{fg:diareal}, and the corresponding
crossed contributions with the gluon in the initial state.
\begin{figure}[hbt]
\SetScale{0.7}
{\unitlength 0.7pt
\begin{picture}(120,130)(-100,-10)

\ArrowLine(0,0)(20,20)
\ArrowLine(20,20)(50,50)
\ArrowLine(50,50)(100,50)
\ArrowLine(100,50)(130,20)
\ArrowLine(130,20)(150,0)
\Photon(0,100)(50,50){3}{5}
\Photon(100,50)(150,100){3}{5}
\Gluon(20,20)(130,20){-3}{9}

\put(-15,98){$\gamma$}
\put(-15,-2){$q$}
\put(70,57){$q$}
\put(70,29){$g$}
\put(155,-2){$q'$}
\put(155,98){$W$}
\put(180,48){$+$}

\end{picture}
\begin{picture}(120,110)(-200,-10)

\ArrowLine(0,0)(50,50)
\ArrowLine(50,50)(75,50)
\ArrowLine(75,50)(100,50)
\ArrowLine(100,50)(125,25)
\ArrowLine(125,25)(150,0)
\Photon(0,100)(50,50){3}{5}
\Photon(100,50)(150,100){3}{5}
\GlueArc(100,50)(25,180,315){3}{5}

\put(-15,98){$\gamma$}
\put(-15,-2){$q$}
\put(70,57){$q$}
\put(80,15){$g$}
\put(155,-2){$q'$}
\put(155,98){$W$}
\put(175,48){$+\cdots$}

\end{picture}}
\caption[]{\label{fg:diavirt} \it Typical diagrams of the virtual corrections
to direct photoproduction of $W$ bosons at HERA.}
\end{figure}
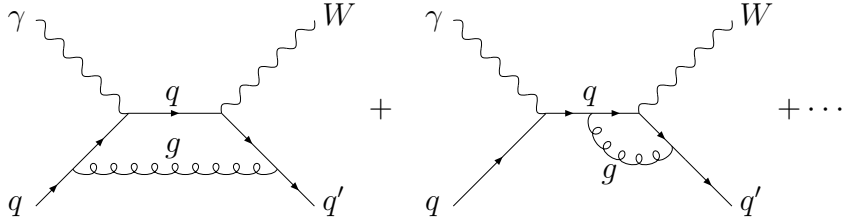
\begin{figure}[hbt]
\SetScale{0.7}
{\unitlength 0.7pt
\begin{picture}(120,110)(-100,-10)

\ArrowLine(0,0)(30,30)
\ArrowLine(30,30)(50,50)
\ArrowLine(50,50)(100,50)
\ArrowLine(100,50)(150,0)
\Photon(0,100)(50,50){3}{5}
\Photon(100,50)(150,100){3}{5}
\Gluon(30,30)(70,0){3}{4}

\put(-15,98){$\gamma$}
\put(-15,-2){$q$}
\put(155,-2){$q'$}
\put(155,98){$W$}
\put(75,-2){$g$}
\put(185,48){$+$}

\end{picture}
\begin{picture}(120,110)(-200,-10)

\ArrowLine(0,0)(50,50)
\ArrowLine(50,50)(75,50)
\ArrowLine(75,50)(100,50)
\ArrowLine(100,50)(150,0)
\Photon(0,100)(50,50){3}{5}
\Photon(100,50)(150,100){3}{5}
\Gluon(75,50)(125,100){3}{7}

\put(-15,98){$\gamma$}
\put(-15,-2){$q$}
\put(155,-2){$q'$}
\put(155,98){$W$}
\put(127,105){$g$}
\put(175,48){$+\cdots$}

\end{picture}}
\caption[]{\label{fg:diareal} \it Typical diagrams of the real corrections
to direct photoproduction of $W$ bosons at HERA.}
\end{figure}
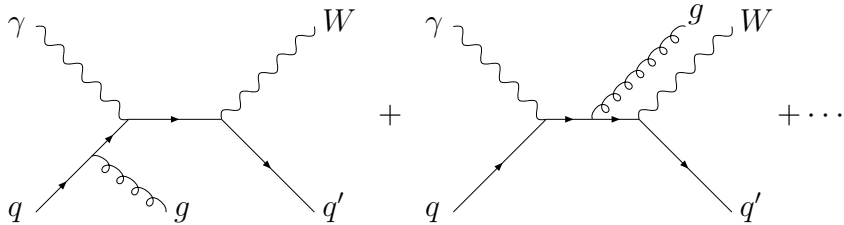

The QCD-corrected result including virtual and real corrections can be
cast into the form
\beq
\frac{d^2\sigma^{dir}_{NLO}}{dp_{TW} dy_W} = 2p_{TW} \sum_{h=q,\bar q,g}
\int_{x_\gamma^-}^1 \frac{dx_\gamma}{x_\gamma} \int_{x_p^-}^1 \frac{dx_p}{x_p}
h_p(x_p,\mu_F^2) f_{\gamma/e} (x_\gamma)
\frac{s^2}{S} \frac{d^2 \hat \sigma^{dir}_{\gamma h}}{dt_1 du_1}
\eeq
with all parameters as defined in
Eqs.\,(\ref{eq:mandelpart}--\ref{eq:mandelhad}) and $h_p$ denoting the
corresponding parton density of the proton. The partonic virtual
corrections are given by
\beq
\frac{d^2 \hat \sigma_{virt}}{dt_1 du_1} = \frac{d \hat \sigma_{virt}}{dt_1}
\delta(s+t_1+u_1+M_W^2)
\eeq
which leads to Eq.\,(\ref{eq:dsiglo}) with $\hat\sigma^{dir}_{LO}$ replaced
by $\hat\sigma_{virt}$ for LO kinematics.

\subsection{Virtual Corrections}
The virtual corrections have been computed via dimensional regularization
in $n=4-2\epsilon$ dimensions. The quarks have been treated as massless
particles. Due to the Ward-Takahashi identities no renormalization is required
after adding all diagrams, since there is no input parameter at LO
which is affected by QCD corrections. The required scalar loop integrals are
taken from Ref.~\cite{loopint}. The final double and single singularities
of the virtual corrections arise from infrared gluon exchange and collinear
singularities due to the gluon and massless quarks,
\beqn
\frac{d\hat\sigma_{virt}}{dt_1} & = & C_{virt} C_F
\frac{\alpha_s}{\pi} \frac{d\hat\sigma_{LO}}{dt_1} \non \\
C_{virt} & = & C_\epsilon \left\{ -\frac{1}{\epsilon^2}
- \frac{1}{\epsilon}\left( \frac{3}{2} + \log\frac{M_W^2}{-t} \right)
+ C_V \right\}
\eeqn
with
\beq
C_\epsilon = \Gamma(1+\epsilon) \left(\frac{4\pi \mu^2}{M_W^2}\right)^\epsilon
\eeq
and a lengthy finite term $C_V$. These singularities are cancelled by
adding the real corrections and the counter terms due to the
renormalization of the parton densities.

As a cross check,
the virtual corrections have also been evaluated by introducing small gluon 
and quark masses $\lambda,m$. This computation has been worked out with the
programs FeynArts \cite{feynarts}, FormCalc \cite{formcalc} and
LoopTools \cite{formcalc}. The renormalization of the small quark mass has
been performed in the $\overline{\rm MS}$ scheme to obtain an
ultraviolet-finite
result. The renormalization, however, does not affect the result in the
limit of the massless gluons and massless quarks, but has to be introduced
to take the massless limit in a consistent way. The infrared and
collinear singularities now appear as logarithms of the gluon and quark
masses in the limit of small masses,
\beq
C'_{virt} = C_\epsilon \left\{ -\log\frac{\lambda^2}{-t}
\log\frac{m^2}{-t} + \frac{1}{2}\log^2\frac{m^2}{-t} - \log\frac{\lambda^2}{-t}
- \frac{1}{2}\log\frac{m^2}{-t} + C'_V \right\}
\eeq
with a different finite term $C'_V$. This result has to be translated into the
$\overline{\rm MS}$ scheme for zero masses. The corresponding piece to be
added to the virtual corrections can be calculated from the results of
Refs.~\cite{neerven,catsey} and it is given by
\beqn
\Delta C_{virt} & = & C_\epsilon \left\{ \log\frac{\lambda^2}{-t}
\log\frac{m^2}{-t} - \frac{1}{2}\log^2\frac{m^2}{-t} + \log\frac{\lambda^2}{-t}
+ \frac{1}{2}\log\frac{m^2}{-t} \right. \non \\
& & \qquad \left. -\frac{1}{\epsilon^2}
- \frac{1}{\epsilon}\left( \frac{3}{2} + \log\frac{M_W^2}{-t} \right) - 2
- \frac{1}{2}\log^2\frac{M_W^2}{-t} - \frac{3}{2}\log\frac{M_W^2}{-t} \right\}
.
\eeqn
After adding this transformation term to the virtual corrections
$C'_{virt}$ of the second approach, both results are in agreement.

\subsection{Real Corrections}
The real corrections have been calculated by means of the massless dipole
subtraction
method introduced in Ref.~\cite{catsey}. The basic idea is to construct
appropriate dipole terms $d\sigma^{sub}$ that include all infrared and
collinear singularities of the real matrix elements, but can be integrated
out analytically up to the LO phase space. The final expression
for the differential cross section is given by
\beq
d \hat \sigma_{NLO} = d\hat \sigma_{LO}
+ [d\hat \sigma^{real}-d\hat \sigma^{sub}]
               + [d\hat \sigma^{virt}+d\bar{\hat \sigma}^{sub}_1]
               + [d\hat \sigma^{part}+d\bar{\hat \sigma}^{sub}_2],
\label{eq:catsey}
\eeq
where $d\bar{\hat \sigma}^{sub}_{1,2}$ are the two parts of the dipole terms
which were integrated out analytically. They cancel the infrared and collinear
divergences of the virtual corrections $d\hat \sigma^{virt}$ and the collinear
singularity of the counter term $d\hat \sigma^{part}$ due to
the renormalization of the parton densities at NLO, respectively. The
NLO parton densities have been defined in the $\overline{\rm MS}$ scheme.
Each of the square brackets in eq.~(\ref{eq:catsey}) is individually finite.
This procedure allows to calculate the real matrix elements in 4 dimensions.

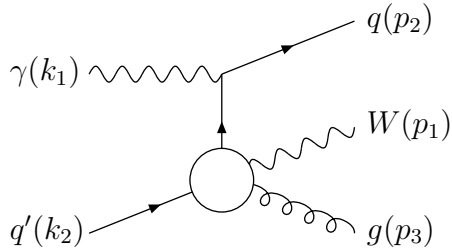
\begin{figure}[hbt]
\begin{picture}(120,100)(-170,10)

\Photon(0,80)(50,80){3}{5}
\ArrowLine(50,80)(100,100)
\ArrowLine(50,40)(50,80)
\ArrowLine(0,20)(50,40)
\Photon(50,40)(100,60){3}{5}
\Gluon(50,40)(100,20){3}{5}
\BCirc(50,40){12}

\put(-30,78){$\gamma (k_1)$}
\put(-30,18){$q'(k_2)$}
\put(105,98){$q (p_2)$}
\put(105,58){$W(p_1)$}
\put(105,18){$g(p_3)$}

\end{picture}
\caption[]{\label{fg:gamqq} \it Generic diagram developing the collinear
singularity between the final state quark and the initial photon.}
\end{figure}

The dipole terms are constructed for each collinear and infrared configuration
arising in the processes. They include the collinear singularities between
final state partons themselves and those between the final state partons
and the initial parton or photon. The collinear singularity related to the
photon is absorbed by the renormalization of the direct part of the resolved
photon densities. There is a subtlety in the numerical implementation of
the dipole term corresponding to collinear $\gamma\to q\bar q$
splitting, see Fig.~\ref{fg:gamqq}.
Although this dipole term does not generate a new spurious singularity, the
numerical integration turns out to be unstable, if the gluon in the final
state is soft. In this case the quark emitted from the photon develops a
large transverse momentum balancing the transverse momentum of the $W$
boson. Since the dipole term has not been constructed to cover this case
numerically, we introduced an upper cut on the product of photon and quark
momenta,
\beq
2k_1p_2 < p_{cut}^2 \ .
\eeq
After introducing this cut into the dipole terms which are subtracted from
the real matrix element, and the analytically integrated one
the numerical integration is stable and independent of this cut.

\section{Results}
We analyze our final results for direct photoproduction of $W$ bosons
plus one jet $W+jet$ as well as the inclusive process $W+X$, i.e.~without
defining jets. We
use the inclusive $k_T$ algorithm for the jet definition with a cone
size $R<1$ \cite{jetalg}. Moreover, we require each detected parton in
the final state
to have transverse energy $E_T > 5$ GeV. One-jet configurations arise in
this framework from 3 different phase space configurations: {\it (i)}
LO kinematics include only one parton in the final state and are
thus one-jet contributions automatically; {\it (ii)} if both partons of the
real corrections are inside the same cone of size $R<1$, they are combined
to one jet; {\it (iii)} if one parton has $E_T<5$ GeV, it is undetected
so that only the other parton is counted as a single jet. The contributions
of these three configurations have to be added to obtain the cross section for
$W+1$ jet production. The residual part of the real corrections is attributed
to $W+2$ jet production which has been added to the 1-jet part for the full
$W+X$ production rate.

We have chosen CTEQ4M
\cite{cteq} and ACFGP \cite{acfgp} parton densities for the proton and the
resolved photon, respectively. The strong coupling constant is taken at NLO
with $\Lambda^{\overline{\rm MS}}_5 = 202$ MeV. At LO we used CTEQ4L parton
densities \cite{cteq}
with LO strong coupling and $\Lambda_5=181$ MeV.
In our numerical analysis we have chosen the following values for the
input parameters: $M_W=80.43$ GeV, $M_Z=91.1876$ GeV,
$\alpha=1/137.03599976$, $G_F=1.16639\times 10^{-5}$ GeV$^{-2}$,
$m_e=0.510998902$ MeV. In the fermionic $Z$ couplings we have introduced
the Weinberg angle $\sin^2\theta_W=1-M_W^2/M_Z^2$,
and the Cabibbo angle between the first two generations has been chosen
as $\sin\theta_c=0.222$ \footnote{In our numerical analysis we neglected
the contribution of initial $b$ quarks.}.

\begin{figure}[hbtp]
\vspace*{0.5cm}

\hspace*{2.0cm}
\begin{turn}{-90}%
\epsfxsize=8cm \epsfbox{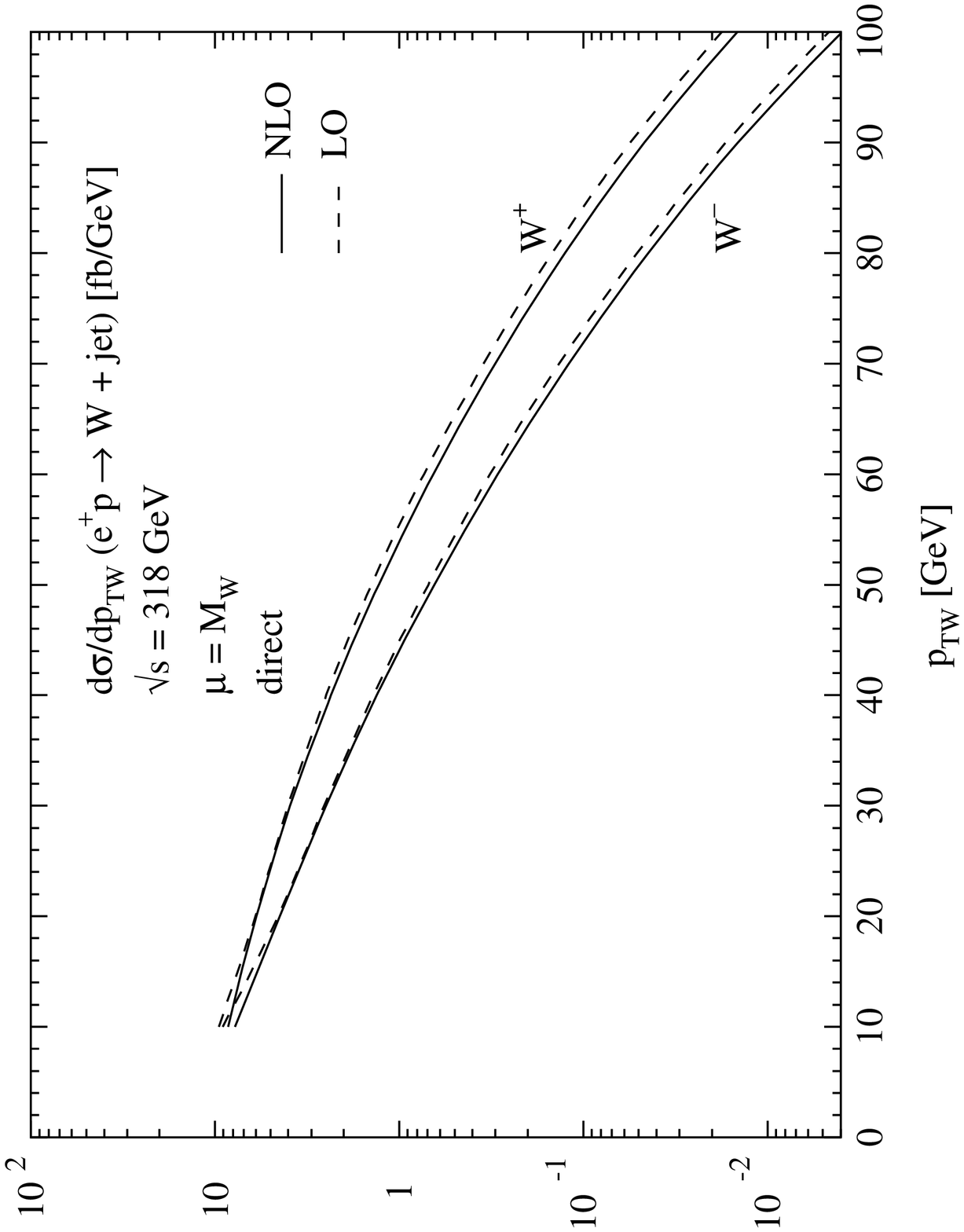}
\end{turn}
\vspace*{0.5cm}

\hspace*{2.0cm}
\begin{turn}{-90}%
\epsfxsize=8cm \epsfbox{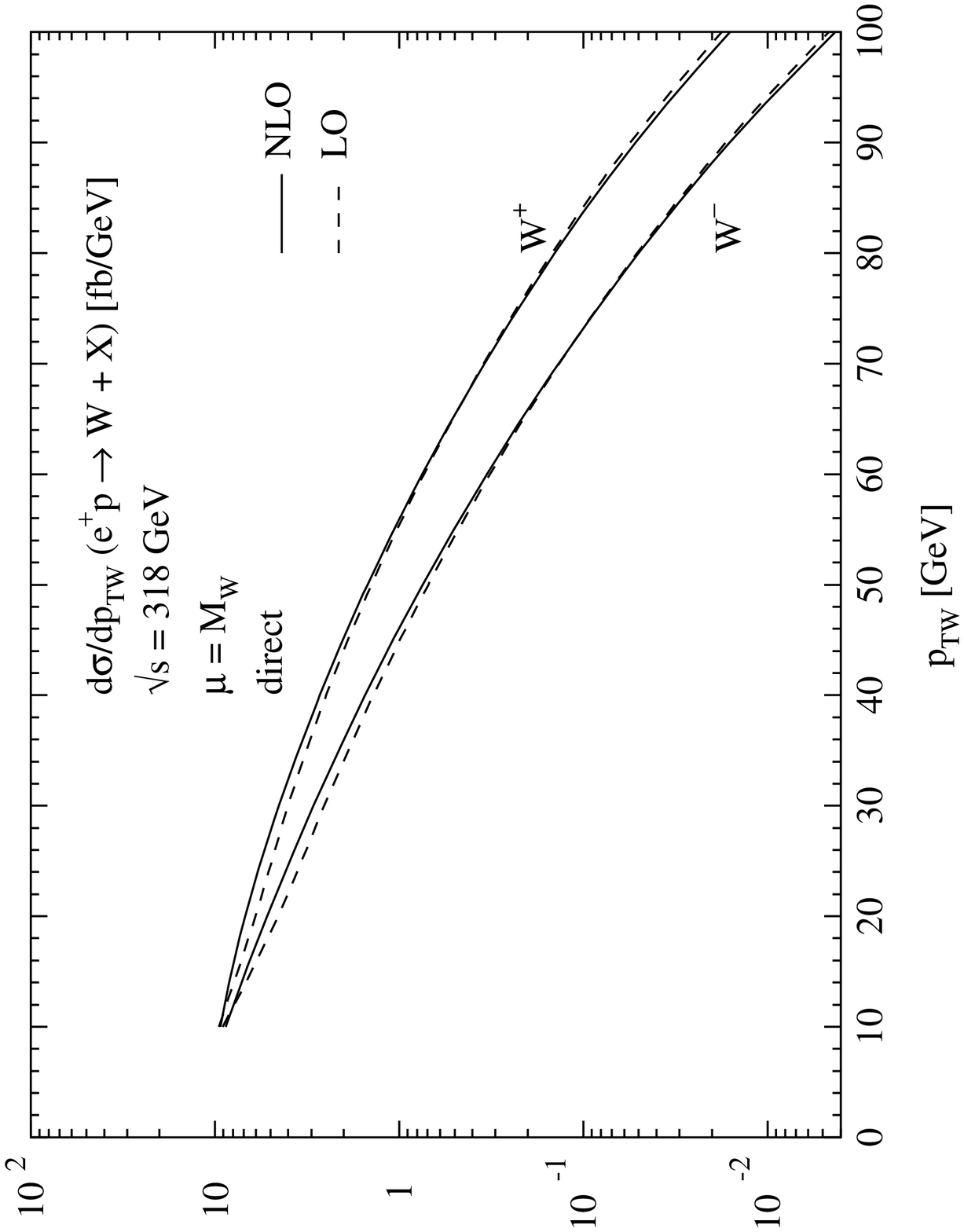}
\end{turn}
\vspace*{0.0cm}

\caption[]{\it \label{fg:pti} Transverse momentum distribution of $W$
bosons at HERA for direct photoproduction. The full curves show the NLO
$p_{TW}$ distributions, while the broken lines exhibit the LO distributions.
The upper plot shows $W+1jet$ production and the lower the total sum of
$W+X$ production.}
\end{figure}
\begin{figure}[hbtp]
\vspace*{0.5cm}

\hspace*{2.0cm}
\begin{turn}{-90}%
\epsfxsize=8cm \epsfbox{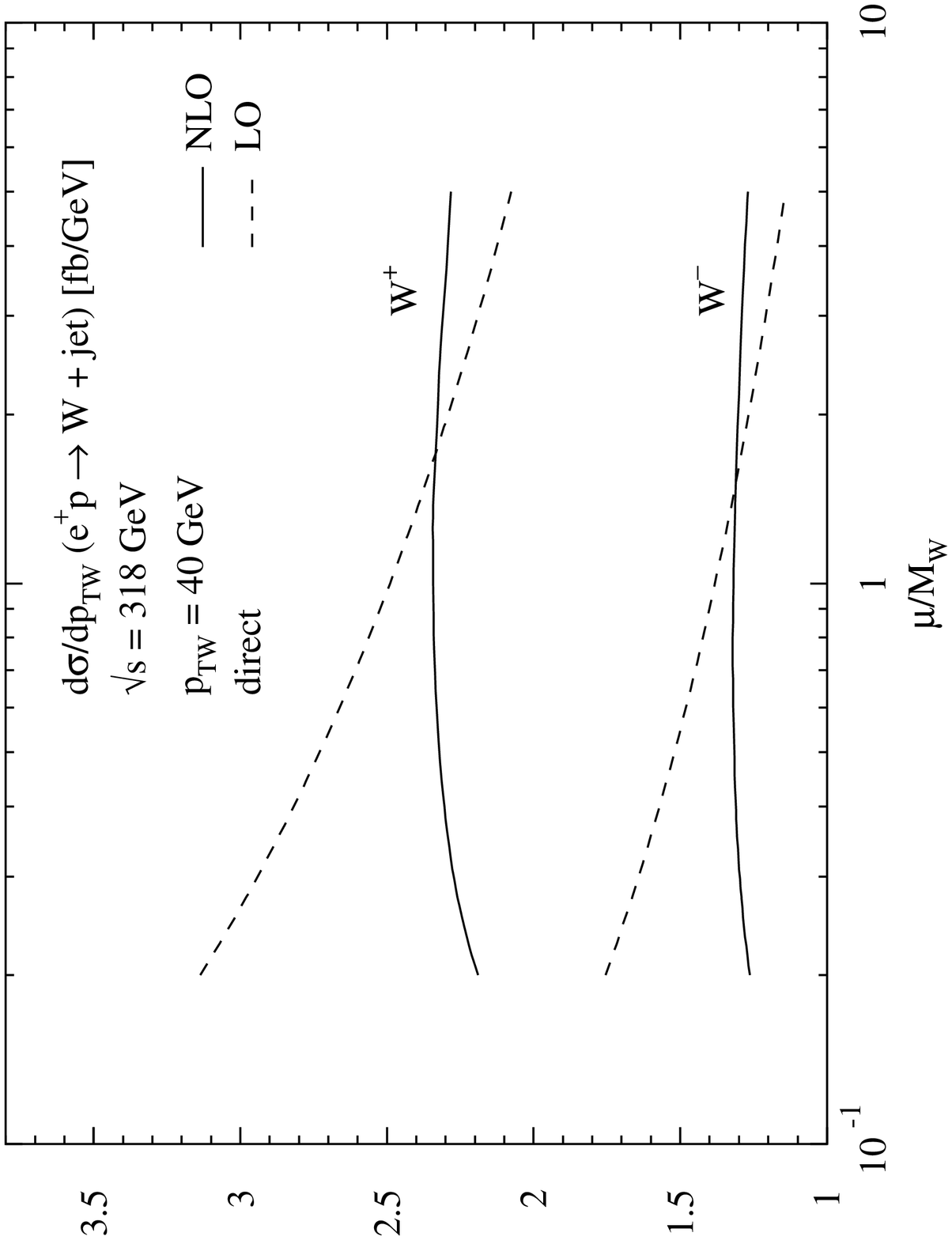}
\end{turn}
\vspace*{0.5cm}

\hspace*{2.0cm}
\begin{turn}{-90}%
\epsfxsize=8cm \epsfbox{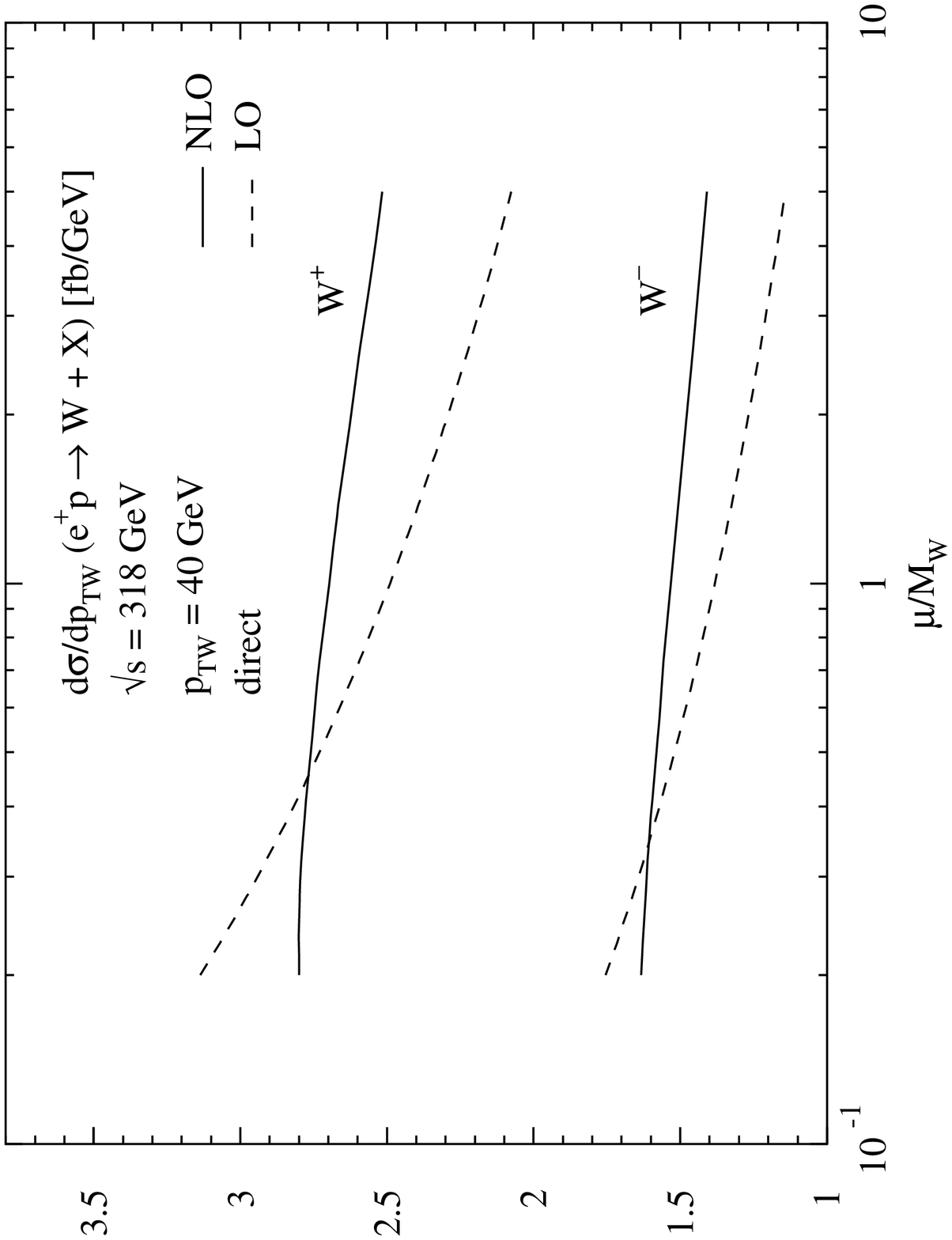}
\end{turn}
\vspace*{-0.0cm}

\caption[]{\label{fg:scale} \it Dependence of the direct contribution
to $W$ production on the
renormalization and factorization scale $\mu = \mu_F = \mu_R = \xi M_W$
for $p_{TW}=40$ GeV.
The full curves represent the NLO predictions and the broken curves the
LO scale dependences.
The upper plot presents $W+1jet$ production and the lower the total sum of
$W+X$ production.}
\end{figure}

Setting $\mu_R = \mu_F = M_W$ for the
renormalization and factorization scales, we present the final results for
the $p_{TW}$ distributions of $W+1$ jet and $W+X$ production in
Fig.~\ref{fg:pti}.
The QCD corrections modify the direct contribution by about $\pm (10-15)\%$ and
they are thus of moderate size.
To estimate the theoretical uncertainties, the
renormalization/factorization scale dependence of the direct
contributions to the processes $e^+ p \to W^\pm + X$
is presented in Fig.~\ref{fg:scale} for HERA conditions. The scale
dependence is significantly smaller, once the NLO corrections are
included. The residual scale dependence is reduced from about 20\% down
to about 5\%. Fig.\,\ref{fg:scale} clearly indicates that the NLO QCD
corrections are accidentially small at the central scale determined by
the $W$ boson mass.
Since the uncertainties of the parton densities are of similar size, the
total theoretical uncertainty can be estimated to be less than about 10\%.

\begin{figure}[hbtp]
\vspace*{0.5cm}

\hspace*{2.0cm}
\begin{turn}{-90}%
\epsfxsize=8cm \epsfbox{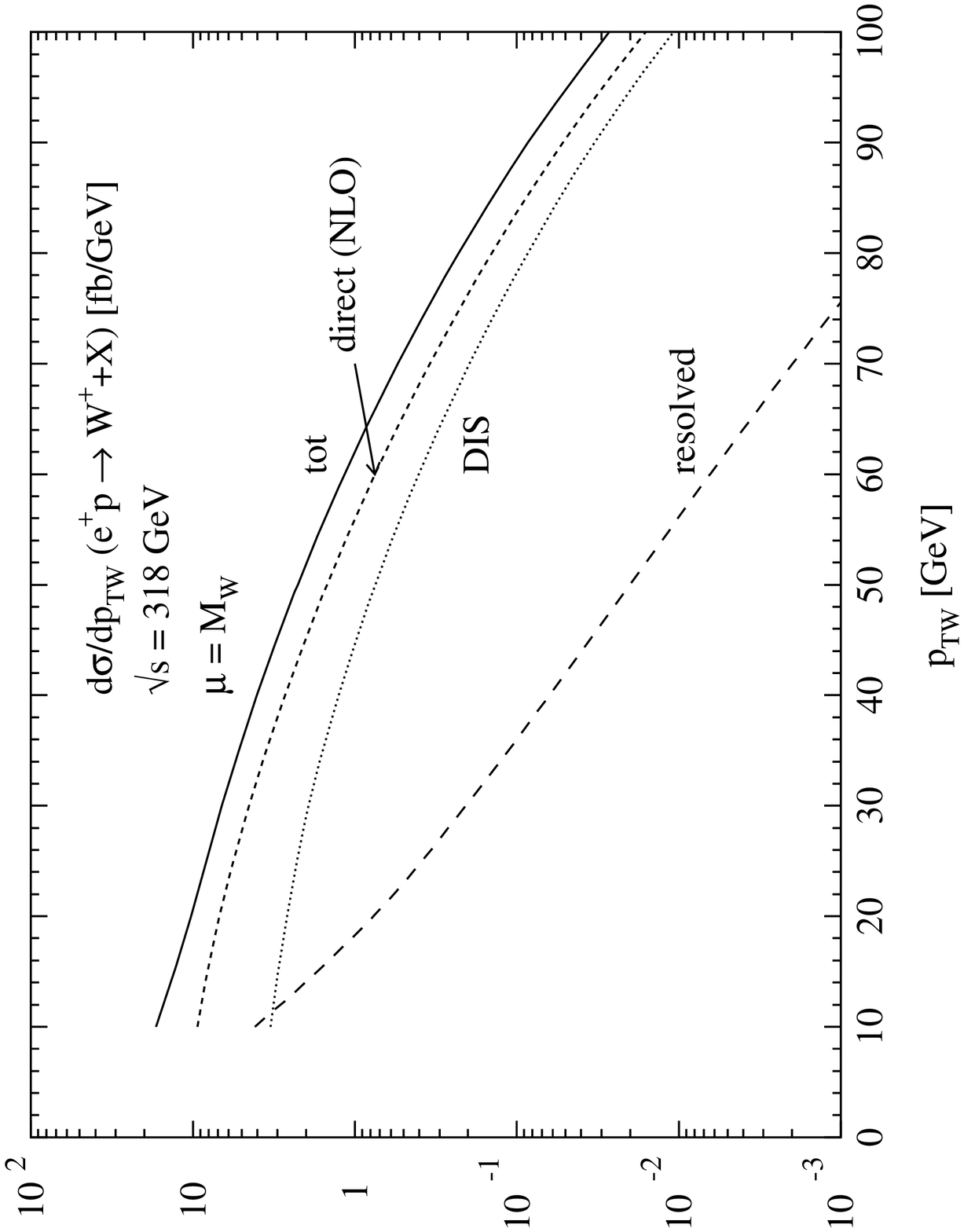}
\end{turn}
\vspace*{0.5cm}

\hspace*{2.0cm}
\begin{turn}{-90}%
\epsfxsize=8cm \epsfbox{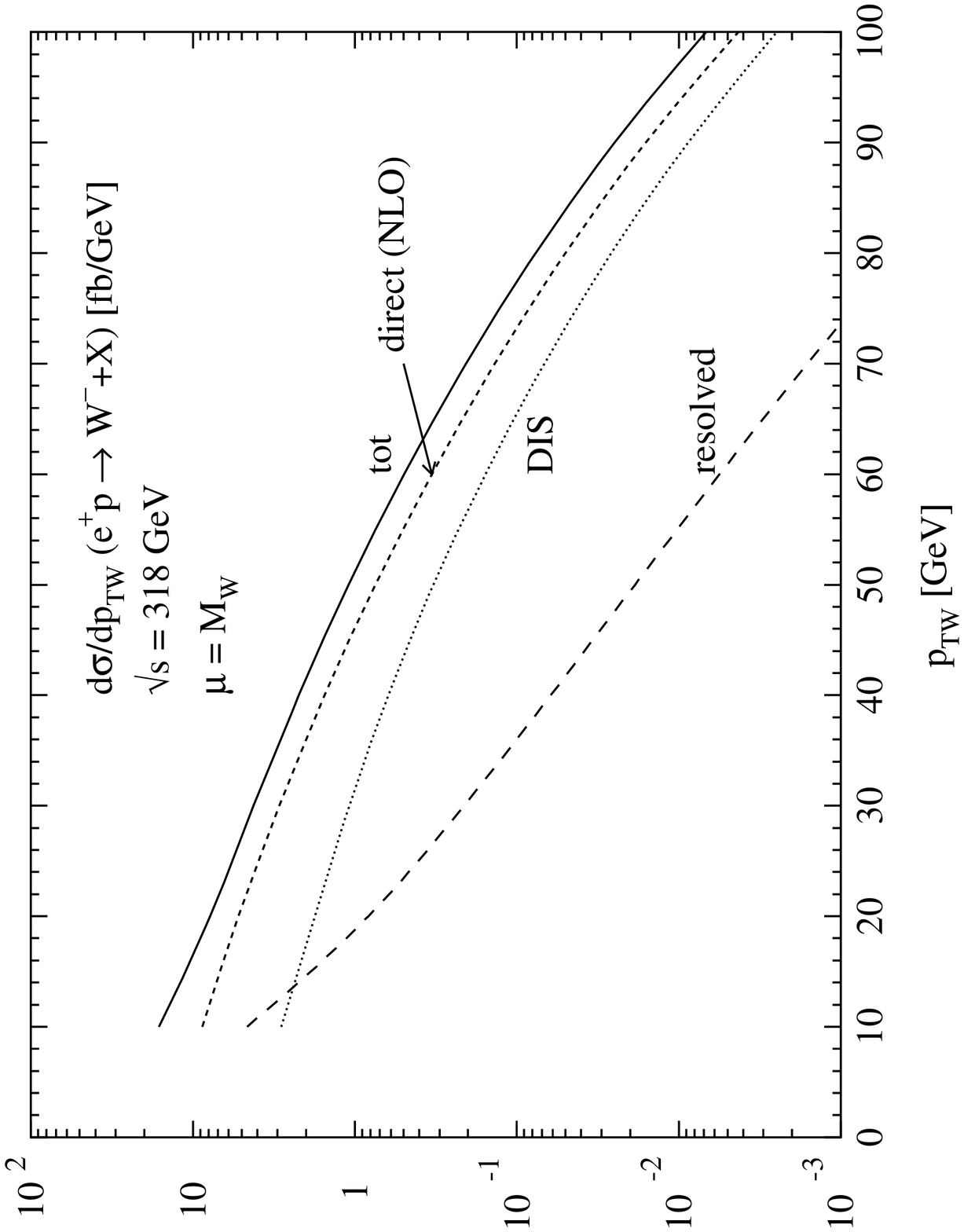}
\end{turn}
\vspace*{0.0cm}

\caption[]{\it \label{fg:ptfull} Transverse momentum distributions of $W$
bosons at HERA. The full curves show the total $p_{TW}$ distributions, while
the broken lines exhibit the individual LO DIS, NLO direct and LO resolved
contributions.
The upper plot is for $W^+$ production and the lower for $W^-$ bosons.}
\end{figure}

In Fig.~\ref{fg:ptfull} we present our final results for the total transverse
momentum distributions of $W^+$ and $W^-$ production at HERA including the NLO
QCD corrections to the direct photoproduction part. The individual
contributions
are also shown. The Figure clearly demonstrates the dominance of the
direct photoproduction part, while the DIS part is smaller but still
relevant. The resolved photoproduction part can safely be neglected for
$p_{TW}$ values larger than about $10-15$ GeV. The rising total sum at $p_{TW}$
values below about 15 GeV signals the breakdown of the pure perturbative
results and underlines the necessity of a soft gluon resummation
to describe $W$ boson production for smaller transverse momenta
which however is beyond the scope of this work.

\begin{figure}[hbtp]
\vspace*{0.5cm}

\hspace*{2.0cm}
\begin{turn}{-90}%
\epsfxsize=8cm \epsfbox{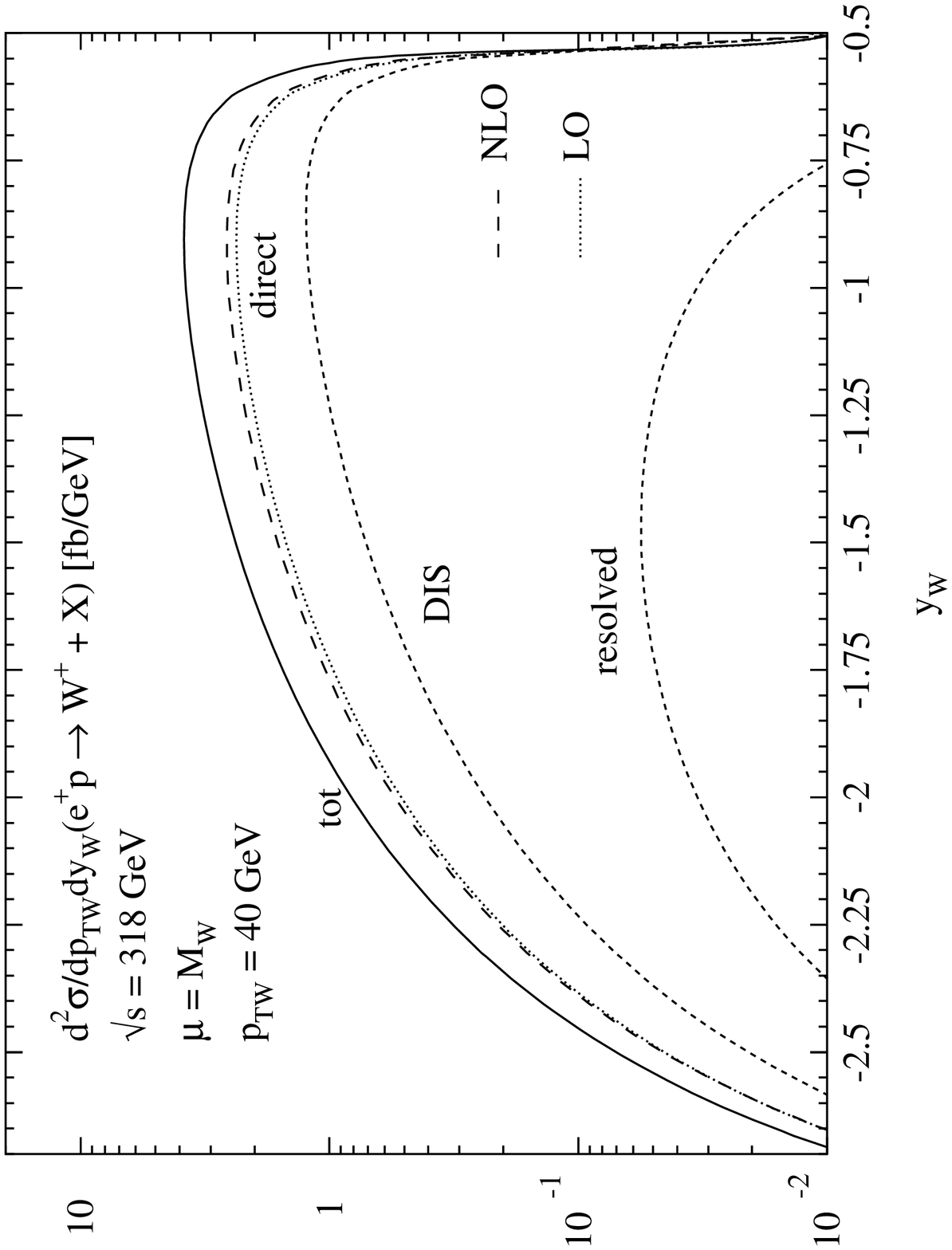}
\end{turn}
\vspace*{0.5cm}

\hspace*{2.0cm}
\begin{turn}{-90}%
\epsfxsize=8cm \epsfbox{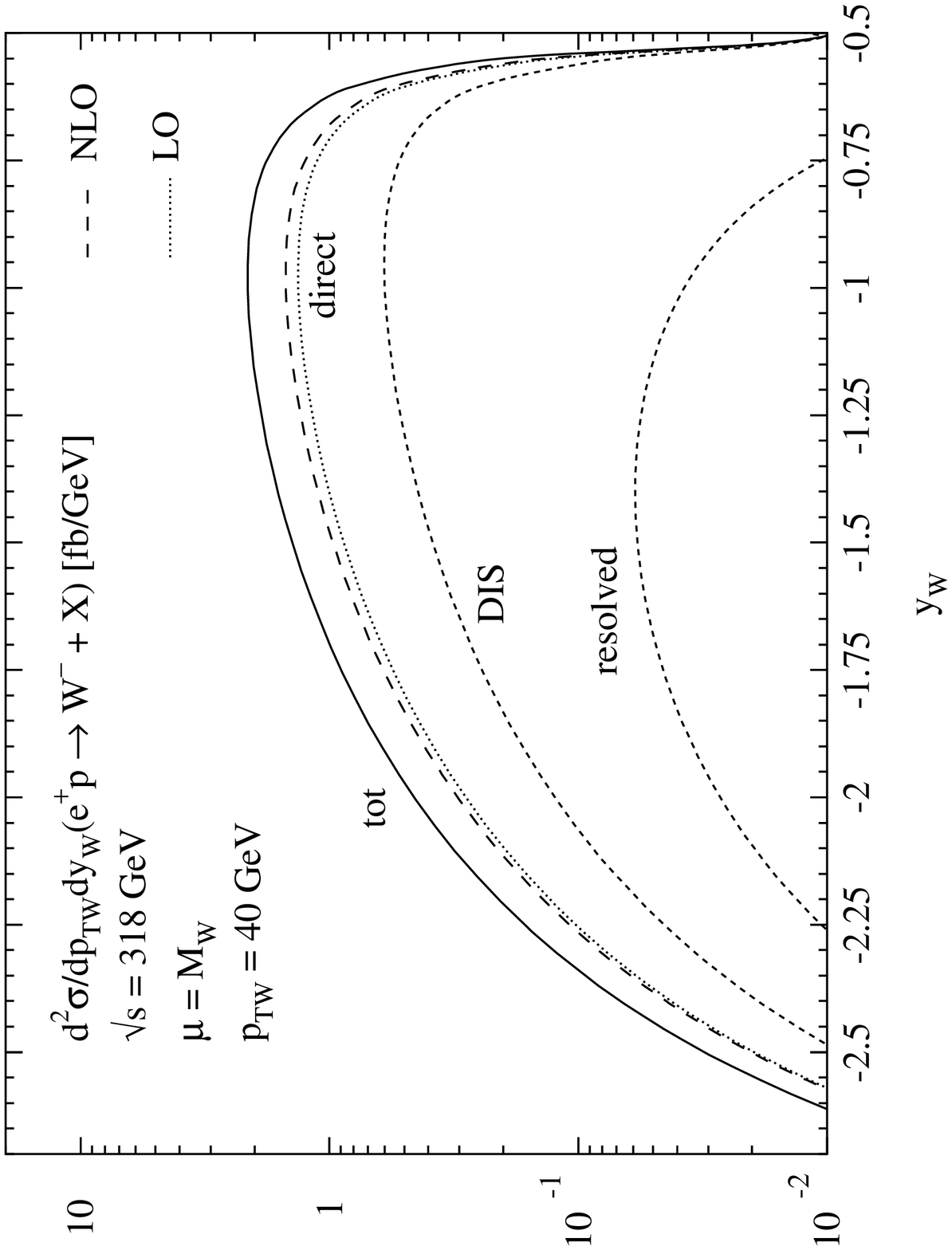}
\end{turn}
\vspace*{0.0cm}

\caption[]{\it \label{fg:yi} Rapidity distribution of $W$ bosons at HERA
at $p_{TW}=40$ GeV. The full curves show the
total distributions including the NLO corrections to the direct contribution,
while the broken lines exhibit the individual parts at LO and the direct
part at NLO. The upper plot presents $W^+$ production and the lower plot
$W^-$ production.}
\end{figure}

The rapidity distribution of $W$ boson production at HERA for
$p_{TW}=40$ GeV is presented in Fig.~\ref{fg:yi} for all individual
parts. The direct photoproduction part is shown at LO and NLO.
As in the case of the transverse momentum
distribution the QCD corrections to the direct contribution amount to less
than about $10-15\%$ and are of moderate size. They hardly change the shape
of the rapidity distribution. It can clearly be inferred from this figure
that there is a preference to produce $W$ bosons at larger values of the
rapidity $y_W$.

In order to allow a comparison of our results with existing analyses at HERA,
we have determined the cross sections for transverse momenta below and
above 25 GeV. In the first case we have used the NLO results of
Refs.~\cite{NRS,spira} for the cross section with $p_{TW}<25$ GeV, while
for $p_{TW}>25$ GeV we have integrated our results including the NLO
corrections to direct photoproduction. We obtain the following values
for the cross section in the case of a positron beam (the numbers in
brackets are the LO values):
\begin{eqnarray}
p_{TW}<25~{\rm GeV} &:& \sigma(W^+) = 0.478~{\rm pb}
                        \quad (0.363~{\rm pb}) \nonumber \\
                    & & \sigma(W^-) = 0.484~{\rm pb}
                        \quad (0.348~{\rm pb}) \nonumber \\
p_{TW}>25~{\rm GeV} &:& \sigma(W^+) = 0.150~{\rm pb}
                        \quad (0.143~{\rm pb}) \nonumber \\
                    & & \sigma(W^-) = 0.084~{\rm pb}
                        \quad (0.079~{\rm pb}) \ .
\end{eqnarray}
No other cuts have been imposed.

\section{Conclusions}
We have presented predictions for $W$ boson production at HERA including
the QCD corrections to the dominant direct photon mechanism at finite
transverse momentum of the $W$ bosons. Working in the conventional $\MS$
scheme we find that the QCD corrections modify the direct contribution by
about $\pm (10-15)\%$ at the nominal renormalization/factorization scale
$\mu_R=\mu_F=M_W$ and are thus of moderate size. In addition, the NLO
corrections reduce the residual
scale dependence of direct photoproduction to a level of about 5\%.  
Taking into account also the uncertainties in the parton densities of the
proton and the photon, the total theoretical uncertainty is estimated to
be about 10\%. However, the QCD corrections to the DIS part are still
unknown. They are not expected to be significantly larger, because they
have to cancel the $Q^2_{max}$ dependence of the NLO direct contribution
and thus have to be of similar size.
Since the QCD corrections are dominated by soft gluon
effects, the shapes of the differential distributions are hardly
affected. Therefore the results obtained in this work cannot explain the
excess of isolated high-energy lepton events observed at the H1
experiment \cite{muon}. \\

\noindent {\bf Acknowledgements.}\\
We would like to thank S.\,Catani, C.\,Diaconu, R.\,Eichler, E.\,Elsen,
M.\,Kuze, A.\,Mehta, P.\,Schleper and P.M.\,Zerwas for useful discussions.
We are grateful to C.\,Diaconu, P.\,Schleper and P.M.\,Zerwas for carefully
reading the manuscript.

\end{document}